\documentclass[preprint,12pt,a4paper]{elsarticle}

\makeatletter
\def\ps@pprintTitle{%
  \let\@oddhead\@empty
  \let\@evenhead\@empty
  \def\@oddfoot{\reset@font\hfil\thepage\hfil}
  \let\@evenfoot\@oddfoot
}
\makeatother

\usepackage{lineno,hyperref}
\modulolinenumbers[5]

\begin{document}

\begin{frontmatter}

\title{Application of the amended Coriolis flowmeter "bubble theory" to sound propagation and attenuation in aerosols and hydrosols}

\author{Nils T. Basse\fnref{myfootnote}}
\address{Trubadurens v\"ag 8, 423 41 Torslanda, Sweden \\ \vspace{1cm} \today}
\fntext[myfootnote]{nils.basse@npb.dk}



\begin{abstract}
The existing viscous and incompressible theory of isothermal sound propagation and attenuation in suspensions considers solid particles which are infinitely viscous. We extend the theory by applying the amended Coriolis flowmeter "bubble theory". Here, the drag force is a function of both the fluid and particle Stokes numbers and the particle-to-fluid ratio of the dynamic viscosity [V.Galindo and G.Gerbeth, A note on the force of an accelerating spherical drop at low-Reynolds number, Phys. Fluids A {\bf 5}, 3290-3292 (1993)]. Aerosol  and hydrosol examples are presented and differences between the original and extended theories are discussed.
\end{abstract}

\end{frontmatter}



\section{Introduction}

When sound propagates through a suspension, the sound speed is modified and the sound is attenuated; this can have important practical implications, e.g. for jet engines and rocket motors \cite{howe_a}. In this paper, we define a suspension to be any combination of particles entrained in a fluid. The particle can be either a fluid or a solid. Specifically, for aerosols (hydrosols), we define the fluid to be air (water), respectively.

The linear theory of isothermal sound propagation and attenuation in a suspension has been presented in \cite{temkin_a,temkin_b} for solid, i.e. infinitely viscous, particles. We will name this theory the "solid particle" (SP) theory.

Another linear theory of suspensions considers the reaction force on an oscillating fluid-filled container due to entrained particles \cite{hemp_a}. This theory was motivated by the need to model two-phase flow in Coriolis flowmeters and is known as the "bubble theory". The bubble theory has been used to model both (i) measurement errors \cite{basse_a} and (ii) damping \cite{basse_b} experienced by Coriolis flowmetering of two-phase flow. The analogy between the bias flow aperture theory \cite{howe_b} and the bubble theory has been explored in \cite{basse_c}. In this paper, we use an amended bubble theory which is named the "viscous particle" (VP) theory.

Both theories consider viscous effects in incompressible fluids. It has been shown that for low frequencies, the assumption of an incompressible fluid is valid for acoustics \cite{temkin_a}. In terms of included physics in the two theories, the main difference is that another drag force \cite{galindo_a} is included in the VP theory; this drag force is a function of both the fluid and particle Stokes numbers and of the particle-to-fluid ratio of the dynamic viscosity. It is important to note that this VP theory is an amended version of the original bubble theory, where we here use \cite{galindo_a} instead of \cite{yang_a} for the drag force on the particle. The reason for this change is that there are errors in \cite{yang_a} which have been corrected in \cite{galindo_a}. For the SP theory, the drag force only depends on the fluid Stokes number - which depends on the dynamic viscosity of the fluid but not of the particle. Thus, the SP theory is a limiting case of the more general VP theory.

We note that a theory exists where the fluid is both viscous and compressible \cite{temkin_c,temkin_a,temkin_b}; for Coriolis flowmeters, a corresponding theory exists where the fluid is inviscid and compressible \cite{hemp_b}. None of these compressible theories take the dynamic viscosity of the particle into account which is the topic of this paper.

The paper is organised as follows: In Section \ref{sec:velrat}, we present theoretical expressions for the particle-to-fluid velocity ratio and apply these to compare aerosols and hydrosols using both theories. We use the same mixture examples to compare sound propagation and attenuation in Section \ref{sec:prop_att} and conclude in Section \ref{sec:conc}.

\section{Particle-to-fluid velocity ratio}
\label{sec:velrat}

The particle-to-fluid velocity ratio is given as:

\begin{equation}
V=\frac{u_p}{u_f},
\end{equation}

\noindent where $u_p$ is the particle velocity and $u_f$ is the fluid velocity.

\subsection{Theories}

\subsubsection{Common assumptions and differences}

For both theories, there are a number of common assumptions:

\begin{itemize}
  \item Both the particles and fluid are considered to be incompressible.
  \item The particles are rigid, i.e. they are spherical and do not deform.
  \item The theories are valid for low Reynolds number.
  \item The drag force for both theories is a function of the fluid Stokes number.
\end{itemize}

Simultaneously, the main conceptual and physical differences are:

\begin{itemize}
  \item The angular frequency $\omega$ has a different physical meaning for the two theories: For the SP theory, it is the acoustic wave frequency and for the VP theory, it is the frequency of the container oscillation. However, mathematically they are completely equivalent.
  \item The drag force constitutes the physical difference between the theories; for the VP theory, the drag force depends on the dynamic viscosity of the particle, which is not the case for the SP theory. The drag force for the VP theory also depends on the particle Stokes number which is not considered in the SP theory.
\end{itemize}

\subsubsection{Solid particle theory}

This section contains the equations of a sphere in an oscillating fluid as described in \cite{temkin_a}.

The particle-to-fluid velocity ratio is:

\begin{equation}
\label{eq:SP_V}
V_{\rm SP}=3 \delta \frac{\beta_f(2\beta_f+3)+3{\rm i} (1+\beta_f)}{2\beta_f^2(2+\delta)+9\beta_f\delta + 9{\rm i}\delta(1+\beta_f)},
\end{equation}

\noindent where $\beta_f$ is the fluid Stokes number and $\delta=\rho_f/\rho_p$. Here, $\rho_f$ is the fluid density and $\rho_p$ is the particle density. The fluid Stokes number is:

\begin{equation}
\label{eq:f_Stokes}
\beta_f = a \sqrt{\frac{\omega \rho_f}{2 \mu_f}},
\end{equation}

\noindent where $a$ is the particle radius and $\mu_f$ is the dynamic viscosity of the fluid.

From Eq. (\ref{eq:SP_V}), the amplitude of the velocity ratio is:

\begin{equation}
|V_{\rm SP}| = 3\delta\sqrt{\frac{4\beta_f^4+12\beta_f^3+18\beta_f^2+18\beta_f+9}{4(2+\delta)^2\beta_f^4+36\delta \beta_f^3(2+\delta)+81\delta^2(2\beta_f^2+2\beta_f+1)}},
\end{equation}

\noindent and the phase angle of the velocity ratio is:

\begin{equation}
\tan \eta_{\rm SP} = -\frac{12(1+\beta_f)\beta_f^2(1-\delta)}{4\beta_f^4(2+\delta)+12\beta_f^3(1+2\delta)+27\delta(2\beta_f^2+2\beta_f+1)},
\end{equation}

\noindent where we have changed sign to follow the bubble theory convention \cite{basse_b} that a positive (negative) $\eta$ means that the particles
are leading (lagging) the fluid, respectively.

\subsubsection{Viscous particle theory}

From \cite{hemp_a}, the particle-to-fluid velocity ratio is:

\begin{equation}
\label{eq:reaction_force}
V_{\rm VP}=1+\frac{4(1-\tau)}{4\tau-(9{\rm i} G/\beta_f^2)},
\end{equation}

\noindent where

\begin{equation}
\label{eq:density_ratio}
\tau=1/\delta=\frac{\rho_p}{\rho_f}
\end{equation}

Here, $G$ is proportional to the drag force on the particle as defined in \cite{galindo_a}:

\begin{equation}
G = 1+\lambda_f+\frac{\lambda_f^2}{9}-\frac{(1+\lambda_f)^2}{3+\lambda_f+\kappa g(\lambda_p)},
\label{eq:big_g}
\end{equation}

\noindent where

\begin{equation}
\label{eq:lambda_f}
\lambda_f=(1+{\rm i})\beta_f,
\end{equation}

\begin{equation}
\label{eq:lambda_p}
\lambda_p=(1+{\rm i})\beta_p,
\end{equation}

\noindent and

\begin{equation}
\label{eq:small_g}
g(\lambda_p) = \frac{\lambda_p (6+\lambda_p^2)-3(2+\lambda_p^2) \tanh \lambda_p}{(3+\lambda_p^2) \tanh \lambda_p -3 \lambda_p}
\end{equation}

The viscosity ratio is:

\begin{equation}
\label{eq:visc_ratio}
\kappa=\frac{\mu_p}{\mu_f},
\end{equation}

\noindent where $\mu_p$ is the dynamic viscosity of the particle.

We have also defined the particle Stokes number:

\begin{equation}
\label{eq:p_Stokes}
\beta_p = a \sqrt{\frac{\omega \rho_p}{2 \mu_p}} = \beta_f \sqrt{\frac{\tau}{\kappa}}
\end{equation}

Note that $G$ is defined using \cite{galindo_a}; in \cite{basse_a,basse_b,basse_c}, it was defined using \cite{yang_a} which contains errors. If $\sqrt{\tau/\kappa}=1$, $G$ as defined in \cite{galindo_a} and \cite{yang_a} is identical, see also Tables \ref{tab:aerosols} and \ref{tab:hydrosols}.

As for the SP theory, we write the amplitude of the velocity ratio:

\begin{equation}
|V_{\rm VP}| = \sqrt{Re(V_{\rm VP})^2+Im(V_{\rm VP})^2},
\end{equation}

\noindent and the phase angle of the velocity ratio:

\begin{equation}
\tan \eta_{\rm VP} = \frac{Im(V_{\rm VP})}{Re(V_{\rm VP})}
\end{equation}

\subsubsection{Comparison between the solid and viscous particle theories}

We can recover the SP theory from the VP theory by applying the "solid sphere limit" \cite{legendre_a}, i.e. $\kappa \to \infty$. Here, Equation (\ref{eq:big_g}) reduces to:

\begin{equation}
G_{\kappa \to \infty} = 1+\lambda_f+\frac{\lambda_f^2}{9}
\label{eq:big_g_lim}
\end{equation}

We combine Equations (\ref{eq:reaction_force}) and (\ref{eq:big_g_lim}) to write:

\begin{equation}
\label{eq:reaction_force_lim}
V_{{\rm VP},\kappa \to \infty}=1+\frac{4(1-\tau)}{4\tau-(9{\rm i} G_{\kappa \to \infty}/\beta_f^2)} = V_{\rm SP}^*,
\end{equation}

\noindent where * is the complex conjugate. The complex conjugate is consistent with the sign change of the phase angle as mentioned above. The terms on the right-hand side of Equation (\ref{eq:big_g_lim}) have the following physical meaning \cite{yang_a}:

\begin{itemize}
  \item 1: Quasisteady Stokes drag 
  \item $\lambda_f$: Basset memory
  \item $\frac{\lambda_f^2}{9}$: Added mass
\end{itemize}

For the "bubble limit" \cite{legendre_a}, i.e. $\kappa \to 0$, the final term on the right-hand side of Equation (\ref{eq:big_g}) has to be included. The physical explanation of this term is that it is a new memory term in addition to the Basset memory term \cite{lawrence_a, yang_a, galindo_a, pozrikidis_a}. The inclusion of the final term in the definition of $G$ leads to modifications of both the real and imaginary part of $G$, which in turn leads to changes of both the amplitude and phase of the velocity ratio. A more detailed discussion can be found in \cite{basse_c}.

\subsection{Examples}

\subsubsection{Aerosols}

\begin{figure}
\centering
\includegraphics[width=6cm]{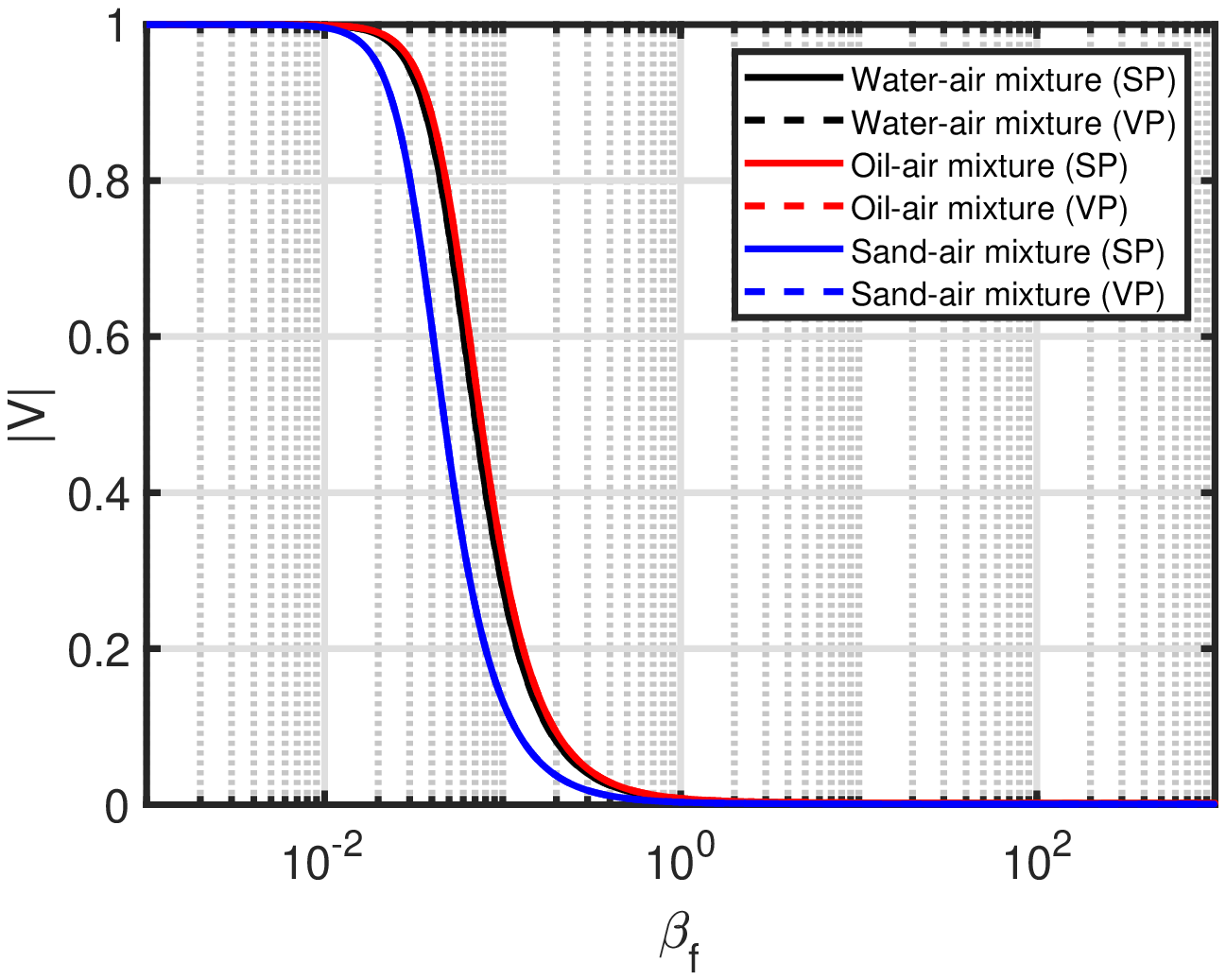}
\includegraphics[width=6cm]{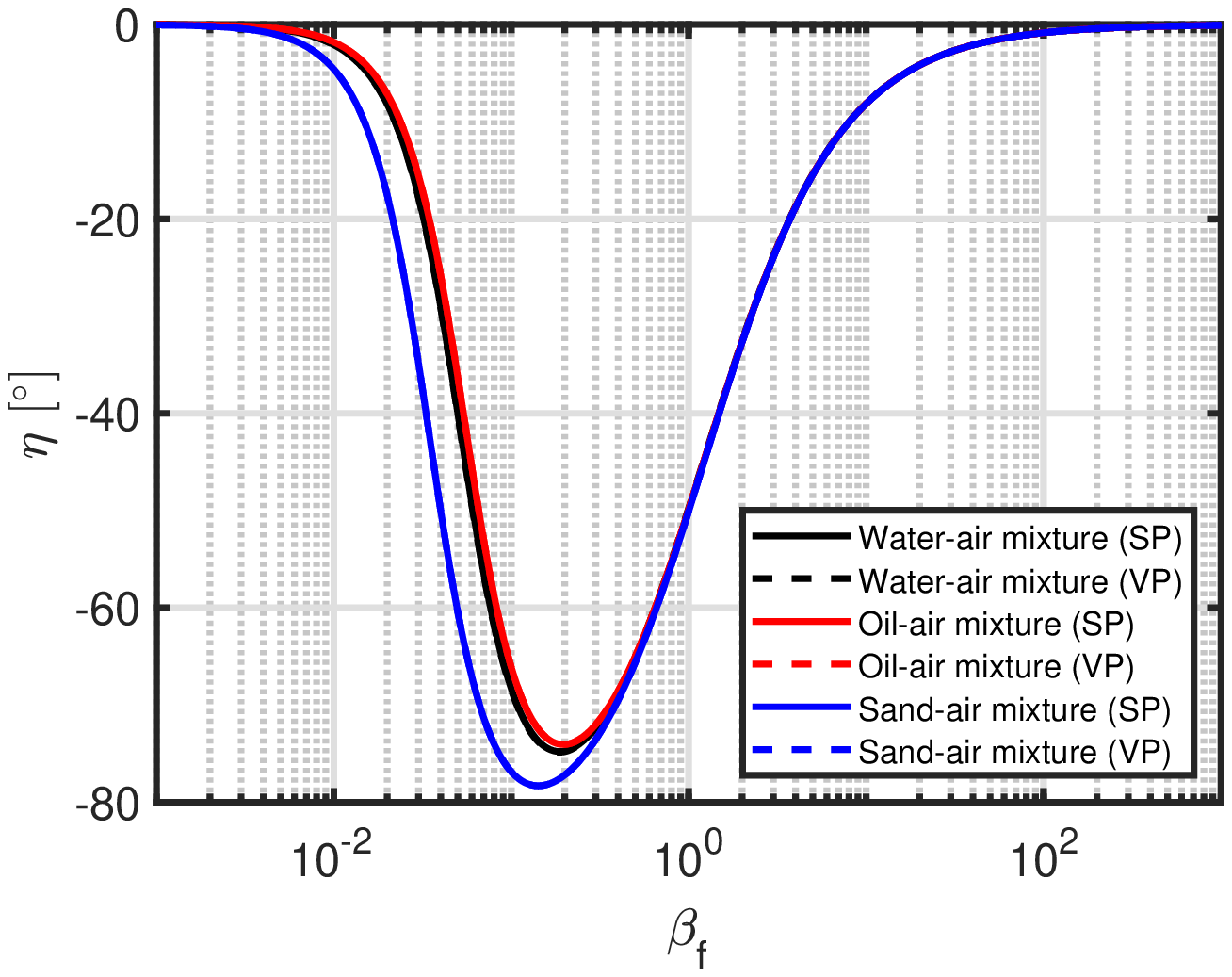}
\caption{Velocity ratio for aerosols: Left: Amplitude, right: Phase. The SP theory is marked by solid lines and the VP theory is marked by dashed lines.}
\label{fig:amp_phase_aerosols}
\end{figure}

Density and viscosity ratios for the aerosols are collected in Table \ref{tab:aerosols}. Corresponding amplitudes and phases of the velocity ratio are shown in Figure \ref{fig:amp_phase_aerosols}.

For small fluid Stokes numbers, the amplitude ratio is close to one, meaning that the particles are moving at the same velocity as the fluid. As the fluid Stokes number increases, the particle velocity decreases with respect to the fluid velocity.

The phase of the velocity ratio becomes negative which means that the particles are lagging the fluid. This is mainly because the density of the particles is much higher than the density of air.

Results from the SP and VP theories are almost identical; thus, inclusion of particle viscosity is not important for aerosols.

\begin{table}[ht]
\caption{\label{tab:aerosols} Density and viscosity ratios for the aerosols treated. Also included is the square root of their ratios.}
\begin{center}
\begin{tabular}{cccc}
 &$\tau$&$\kappa$&$\sqrt{\tau/\kappa}$\\
\hline
Water-air mixture& 831.7  & 50              & 4.1       \\
Oil-air mixture  & 723.3  & 2500            & 0.5       \\
Sand-air mixture & 1833.3 & 5e16 ($\infty$) & 1.9e-7 (0)\\
\end{tabular}
\end{center}
\end{table}

\subsubsection{Hydrosols}

\begin{figure}
\centering
\includegraphics[width=6cm]{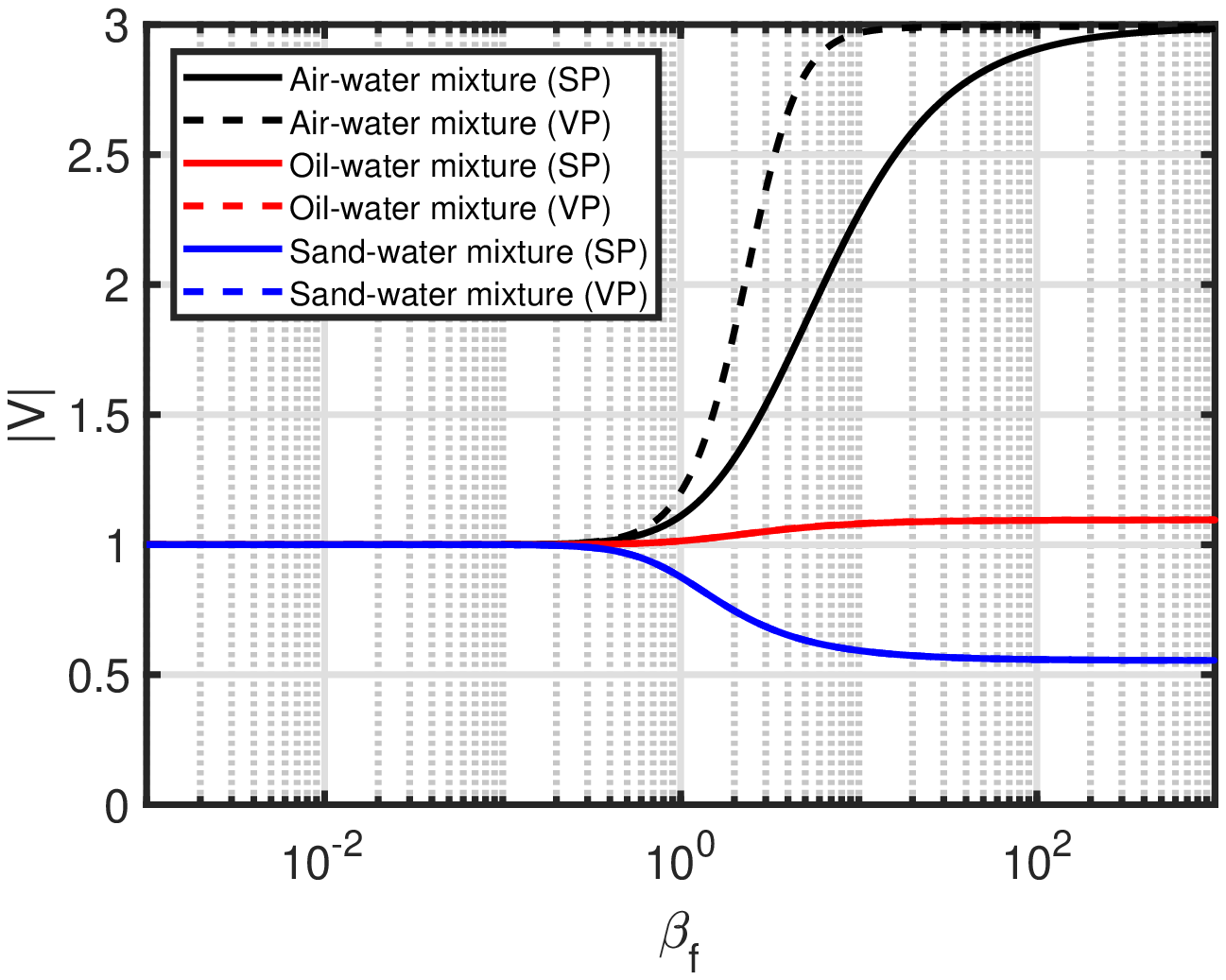}
\includegraphics[width=6cm]{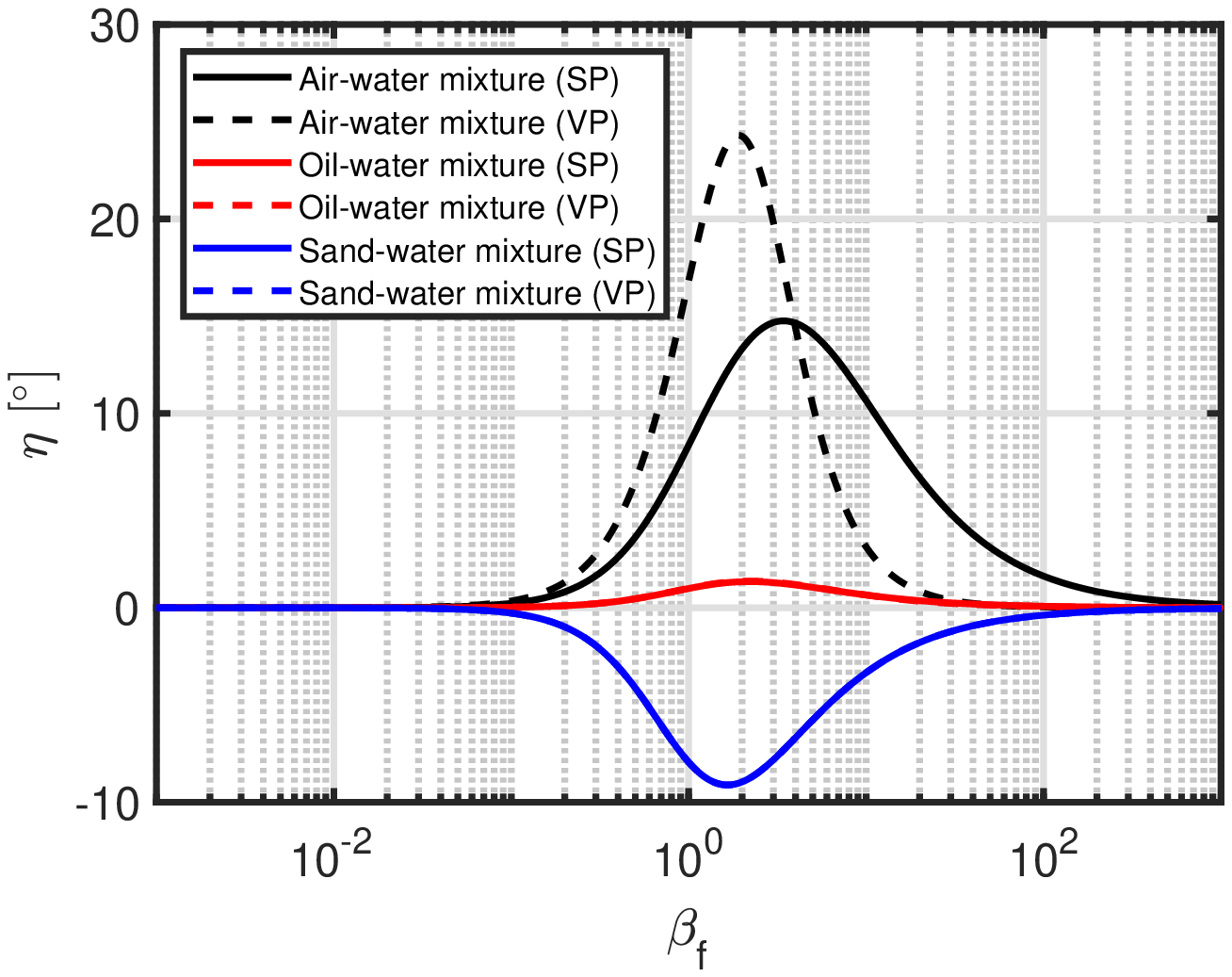}
\caption{Velocity ratio for hydrosols: Left: Amplitude, right: Phase. The SP theory is marked by solid lines and the VP theory is marked by dashed lines.}
\label{fig:amp_phase_hydrosols}
\end{figure}

Density and viscosity ratios for the hydrosols selected are collected in Table \ref{tab:hydrosols}. Corresponding amplitudes and phases of the velocity ratio are shown in Figure \ref{fig:amp_phase_hydrosols}.

\begin{table}[ht]
\caption{\label{tab:hydrosols} Density and viscosity ratios for the hydrosols treated. Also included is the square root of their ratios.}
\begin{center}
\begin{tabular}{cccc}
 &$\tau$&$\kappa$&$\sqrt{\tau/\kappa}$\\
\hline
Air-water mixture & 1.2e-3 & 2e-2               & 0.2       \\
Oil-water mixture & 0.87   & 50                 & 0.1       \\
Sand-water mixture& 2.2    & 1e15 ($\infty$)    & 4.7e-8 (0)\\
\end{tabular}
\end{center}
\end{table}

As for aerosols, the amplitude of the velocity ratio is approximately one for small fluid Stokes numbers; however, it can become both larger and smaller than one for large fluid Stokes numbers. This is mainly a density effect, but as we see for the air-water mixture, there is also an additional effect due to the particle viscosity which is not captured by the SP theory.

The two theories also differ for the phase of the velocity ratio of the air-water mixture, both in the position and maximum value of the peak; the phase is positive, meaning that the particles are leading the fluid.

For the oil-water and sand-water mixtures, the results from the SP and VP theories are almost identical.

What is different for the air-water case is the small value of $\kappa$, see Table \ref{tab:hydrosols}.

\section{Propagation and attenuation of sound}
\label{sec:prop_att}

\subsection{Theory}

\subsubsection{Exact expression}

We present the dispersion relation following analysis as presented in Section 9.4 of \cite{temkin_b}, but only keeping the force source and disregarding the volume and heat sources:

\begin{equation}
\label{eq:1_disp_rel}
\frac{k^2}{k_0^2} = 1 + \phi_{v} ( \tau V -1 ),
\end{equation}

\noindent where $k$ is the complex wavenumber, $k_0$ is the equilibrium wavenumber and $\phi_{v}$ is the volumetric particle fraction. Separating this into real and imaginary parts:

\begin{equation}
\label{eq:X_one}
X(\omega) = Re \left( \frac{k^2}{k_0^2} \right) = 1 + \phi_v \left( \tau Re(V) -1 \right)
\end{equation}

\begin{equation}
\label{eq:Y_one}
Y(\omega) = Im \left( \frac{k^2}{k_0^2} \right) = \phi_{v} \tau Im(V)
\end{equation}

To obtain phase velocity and attenuation, we note that the wavenumber ratio can also be expressed as:

\begin{equation}
\label{eq:2_disp_rel}
\frac{k^2}{k_0^2} = \frac{c_{sf}^2}{c_s^2(\omega)} - \hat{\alpha}^2 + 2 {\rm i} \hat{\alpha} \frac{c_{sf}}{c_s(\omega)},
\end{equation}

\noindent where $c_{sf}$ is the isentropic fluid sound speed, $c_s(\omega)$ is the nonequilibrium isentropic sound speed, $\hat{\alpha} = \alpha c_{sf} / \omega$ is the nondimensional attenuation based on $c_{sf}$ and $\alpha$ is the attenuation coefficient.

As for Equation (\ref{eq:1_disp_rel}), Equation (\ref{eq:2_disp_rel}) can also be separated into real and imaginary parts:

\begin{equation}
\label{eq:X_two}
X(\omega) = Re \left( \frac{k^2}{k_0^2} \right) = \frac{c_{sf}^2}{c_s^2(\omega)} - \hat{\alpha}^2 
\end{equation}

\begin{equation}
\label{eq:Y_two}
Y(\omega) = Im \left( \frac{k^2}{k_0^2} \right) = 2 \hat{\alpha} \frac{c_{sf}}{c_s(\omega)}
\end{equation}

These equations can be solved for the sound speed and attenuation:

\begin{equation}
\frac{c_{sf}^2}{c_s^2(\omega)} = \frac{1}{2} X \left[ 1+\sqrt{1+(Y/X)^2} \right]
\end{equation}

\begin{equation}
\label{eq:alpha_hat}
\hat{\alpha} = \sqrt{X/2} \left[ \sqrt{1+(Y/X)^2} -1 \right]^{1/2}
\end{equation}

We note that there is a typo in the equation for $c_{sf}^2/c_s^2(\omega)$ in \cite{temkin_b} (Equation (9.4.19a)).

\subsubsection{Simplified expression}

For low damping, i.e. $\hat{\alpha} \ll 1$, we can use Equation (\ref{eq:X_two}) to write:

\begin{equation}
\frac{c_{sf}^2}{c_s^2(\omega)} \approx X
\end{equation}

\noindent and use Equations (\ref{eq:Y_two}) and (\ref{eq:Y_one}) to write:

\begin{equation}
\hat{\alpha} = \frac{1}{2} \frac{|Y|}{\sqrt{X}} \approx \frac{1}{2} |Y| = \frac{1}{2} \phi_v \tau |Im(V)|,
\end{equation}

\noindent where we have assumed that $\sqrt{X} \approx 1$.

For this case we can write the scaled attenuation as:

\begin{equation}
\hat{\alpha}/\phi_v \approx \frac{1}{2} \tau |Im(V)|
\end{equation}

\subsection{Examples}

For all examples in this section, we plot both the exact and simplified expressions for the normalized sound speed and the nondimensional attenuation.

Also, we note that the volumetric particle fraction $\phi_{v}$ is set to 0.01 (1\%) for the examples.

\subsubsection{Aerosols}

\begin{figure}
\centering
\includegraphics[width=6cm]{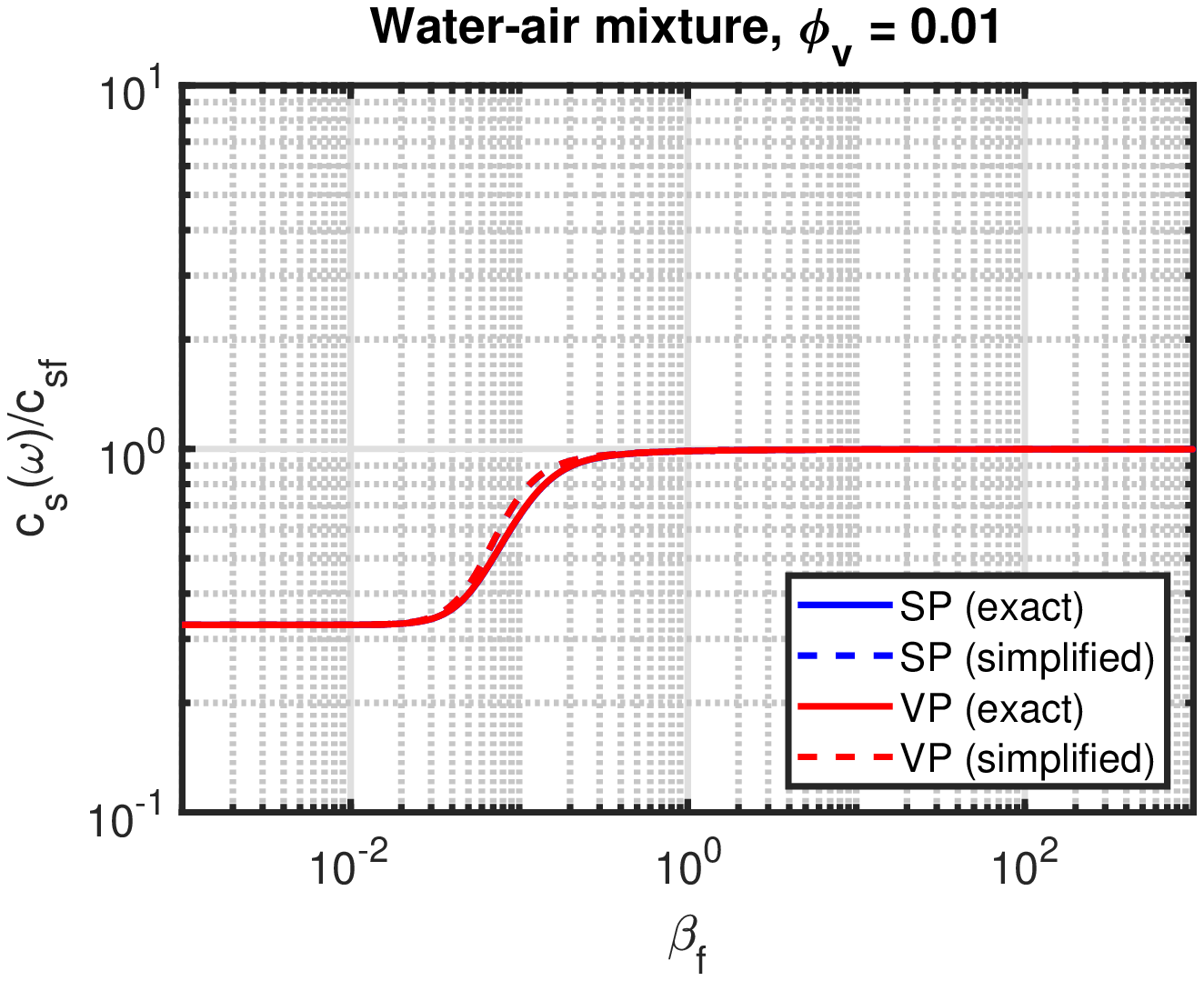}
\includegraphics[width=6cm]{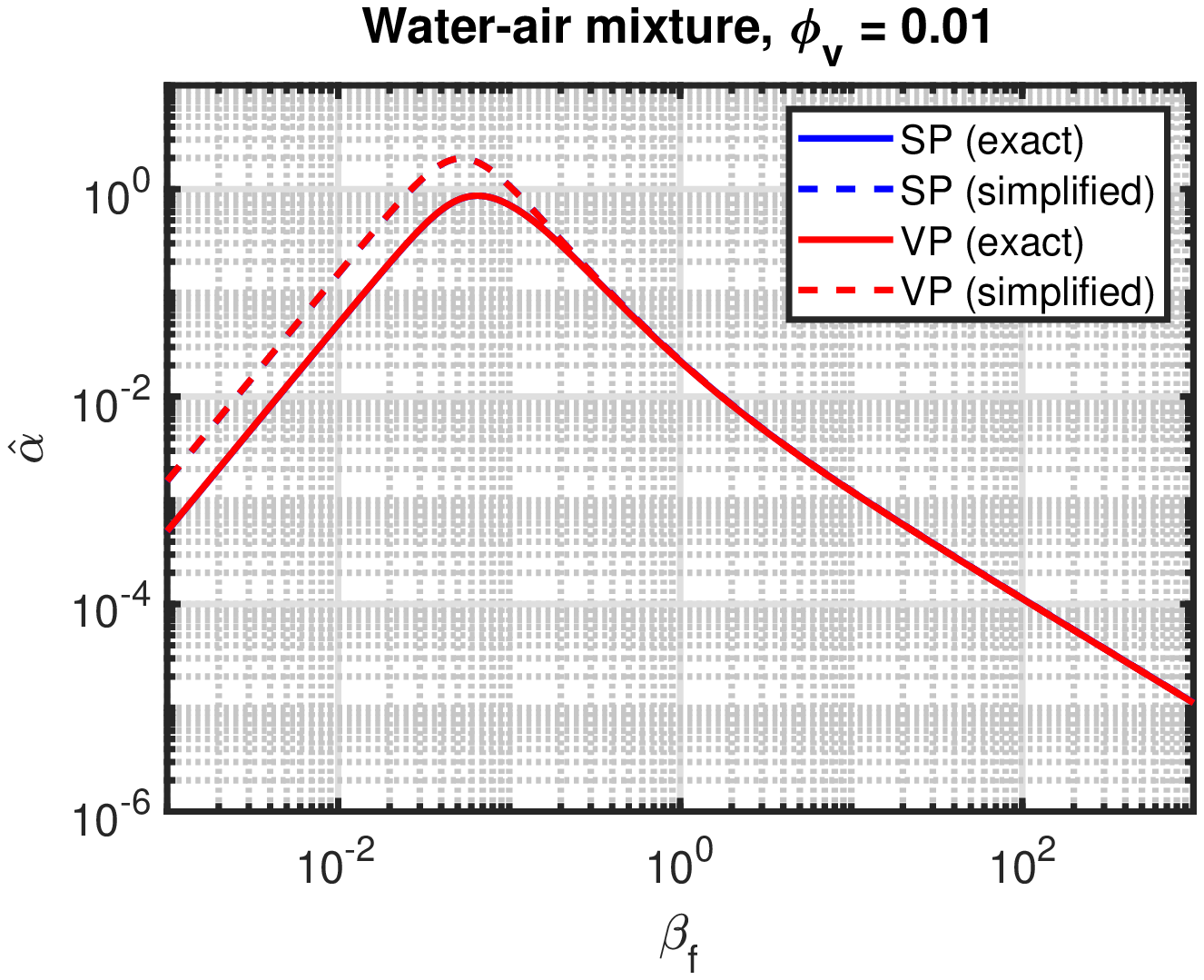}
\caption{Water-air mixture, left: Normalized sound speed, right: Nondimensional attenuation. Exact expressions are marked by solid lines and simplified expressions are marked by dashed lines.}
\label{fig:water_aerosol}
\end{figure}

\begin{figure}
\centering
\includegraphics[width=6cm]{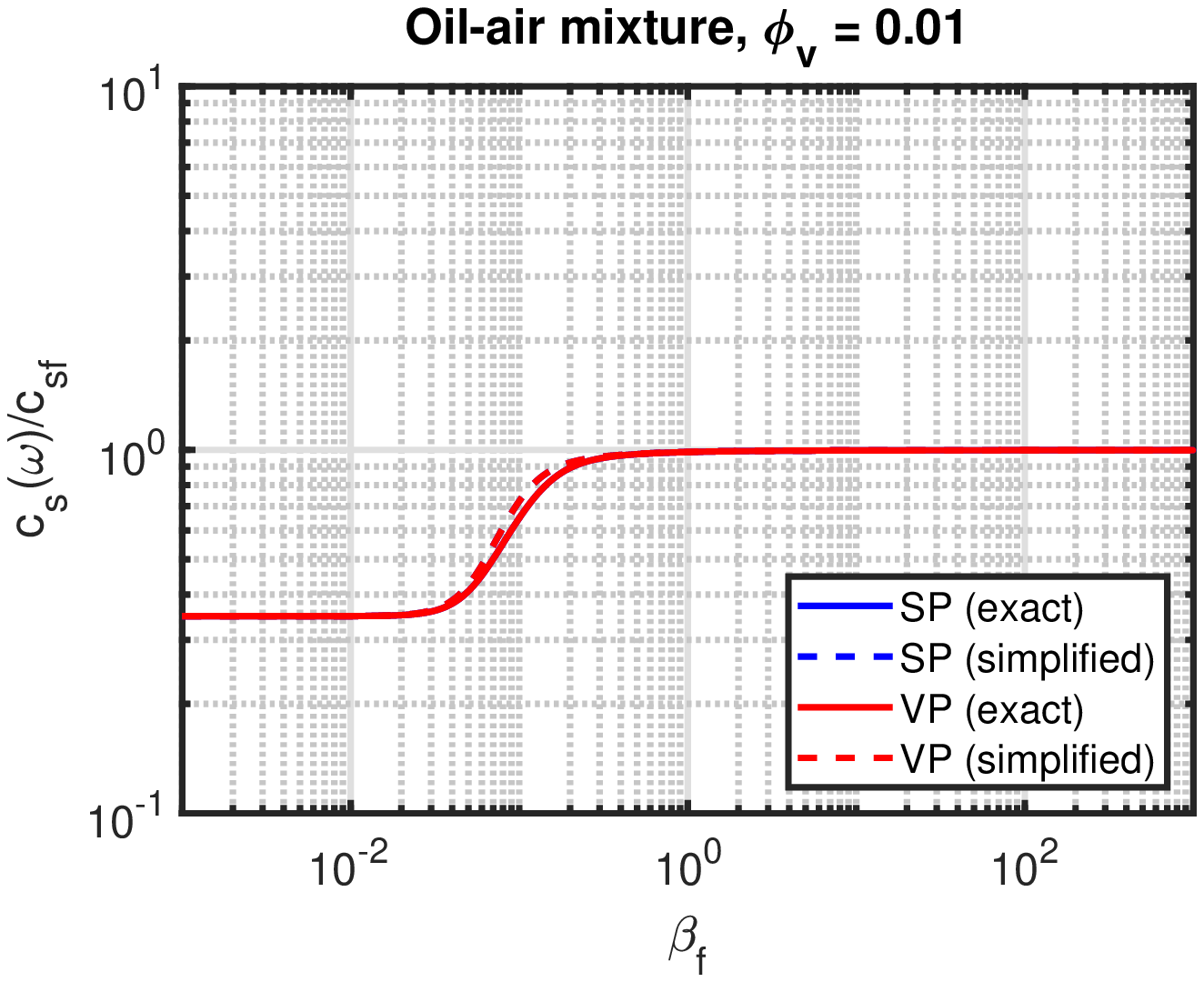}
\includegraphics[width=6cm]{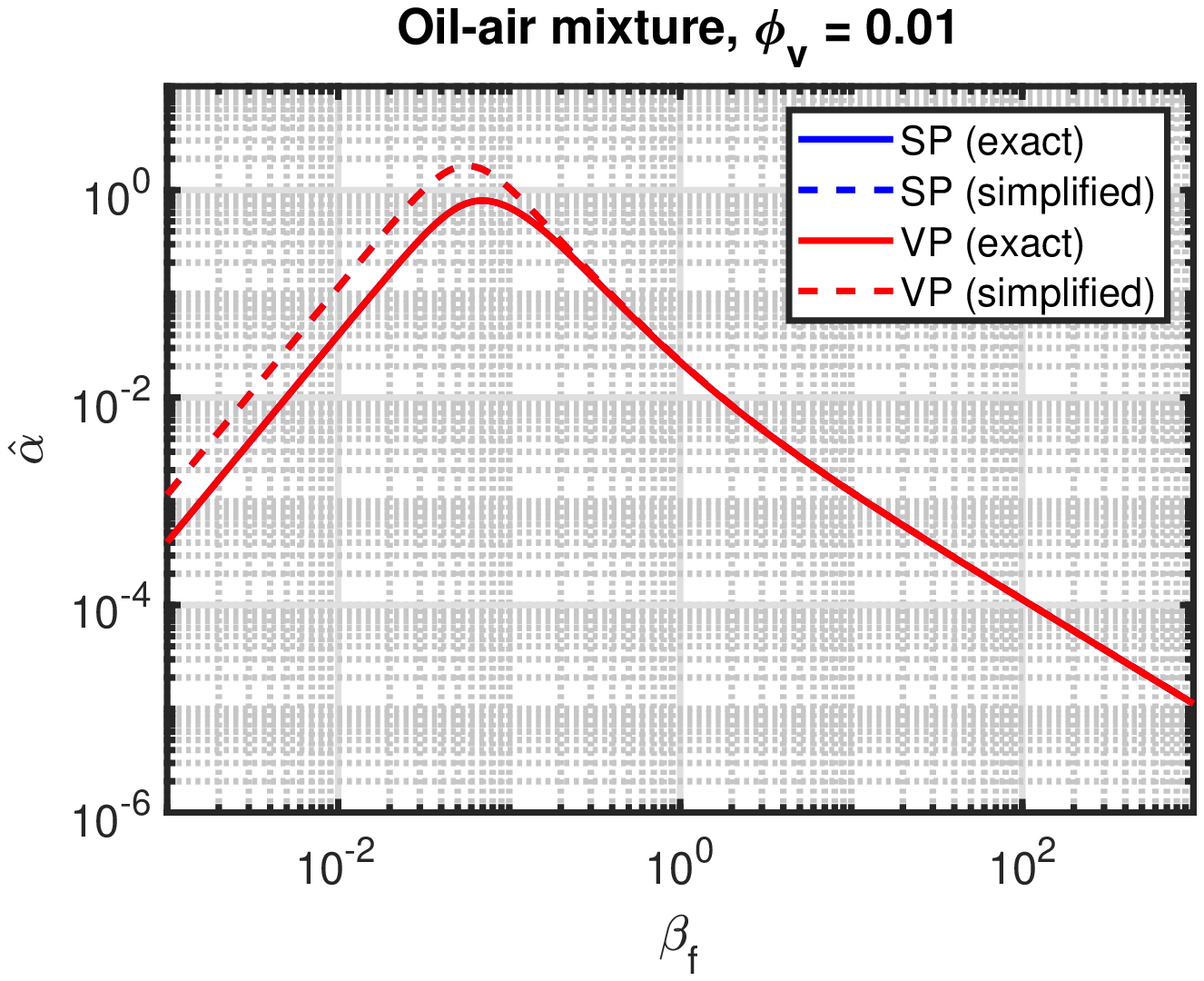}
\caption{Oil-air mixture, left: Normalized sound speed, right: Nondimensional attenuation. Exact expressions are marked by solid lines and simplified expressions are marked by dashed lines.}
\label{fig:oil_aerosol}
\end{figure}

\begin{figure}
\centering
\includegraphics[width=6cm]{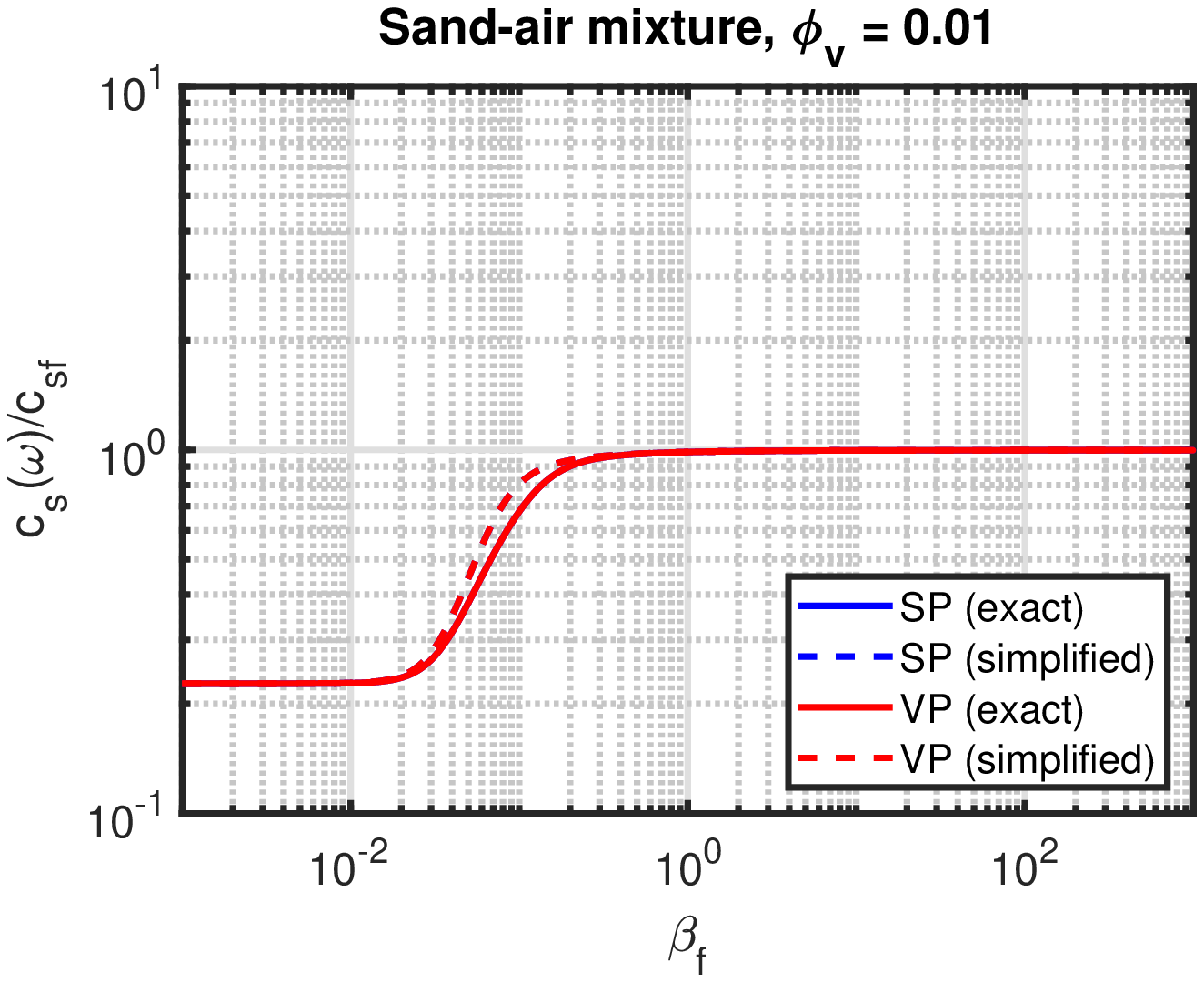}
\includegraphics[width=6cm]{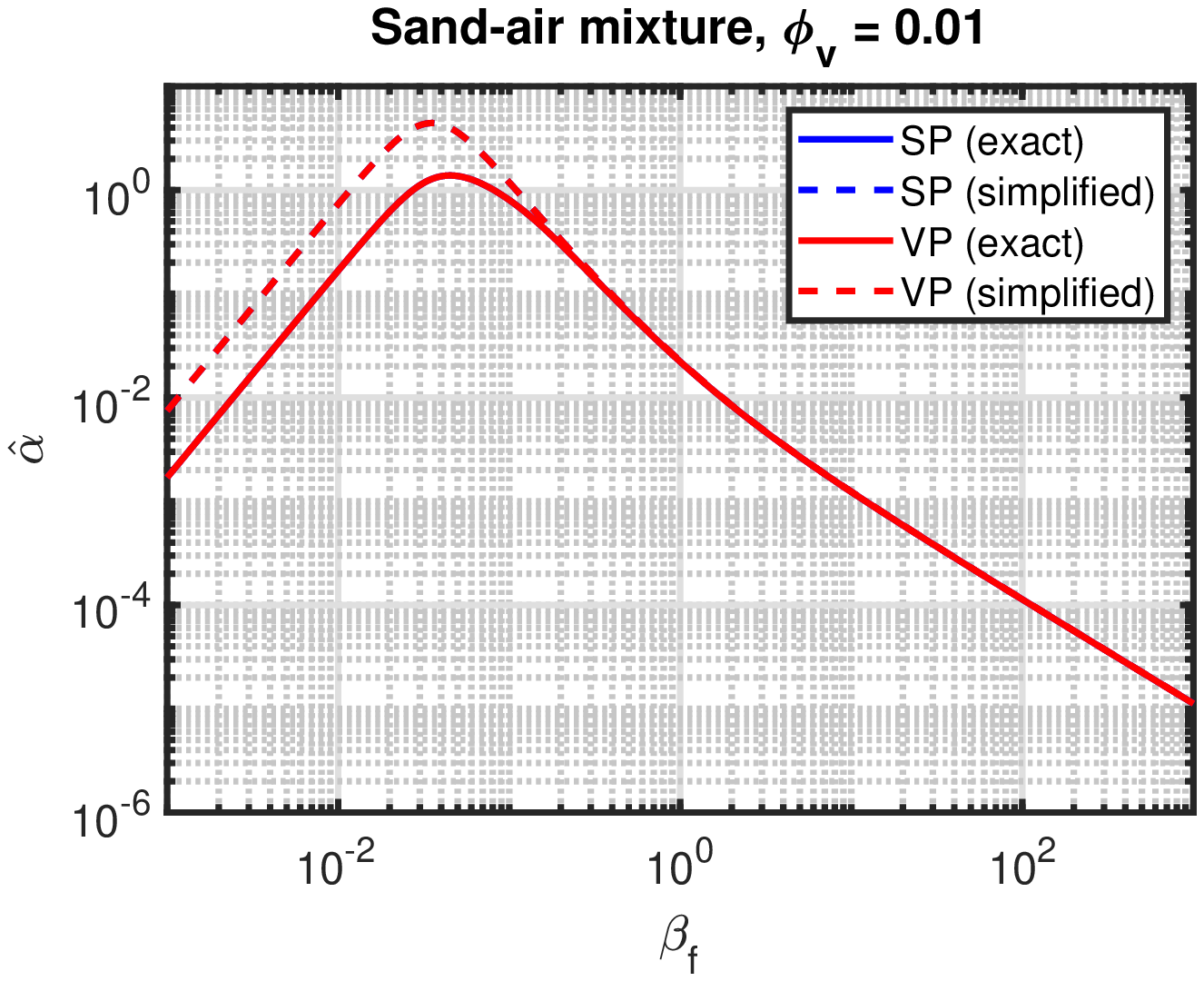}
\caption{Sand-air mixture, left: Normalized sound speed, right: Nondimensional attenuation. Exact expressions are marked by solid lines and simplified expressions are marked by dashed lines.}
\label{fig:sand_aerosol}
\end{figure}

For the aerosols presented in Figures \ref{fig:water_aerosol}-\ref{fig:sand_aerosol}, the sound speed reduces significantly for small fluid Stokes numbers, with ratios in the range of 0.2-0.4.

The peak nondimensional attenuation is of order one, meaning that the simplified expression is not accurate for small fluid Stokes numbers; here, the exact expression should be employed.

Generally, the SP and VP theories agree well for all three aerosol examples.

\subsubsection{Hydrosols}

\begin{figure}
\centering
\includegraphics[width=6cm]{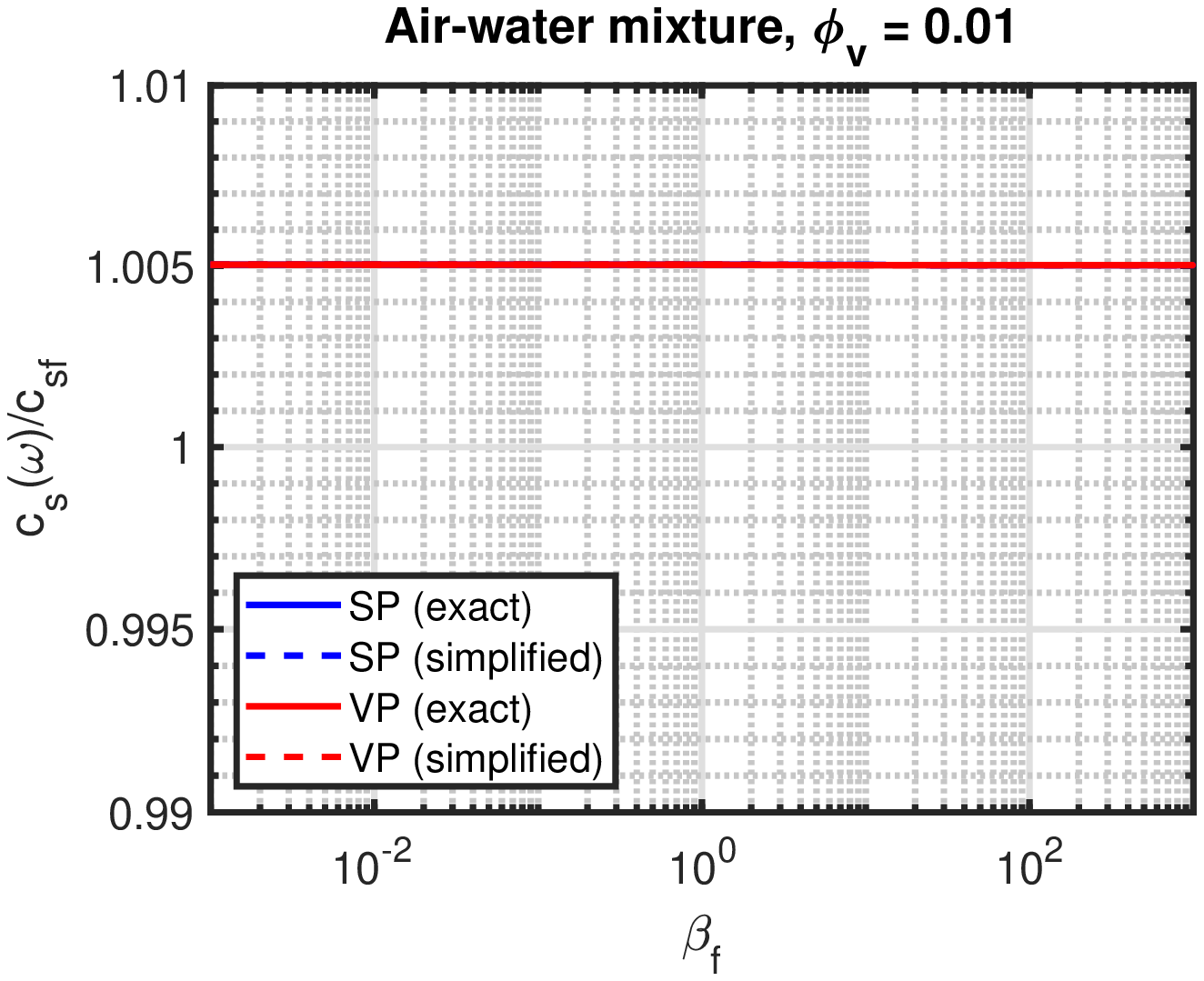}
\includegraphics[width=6cm]{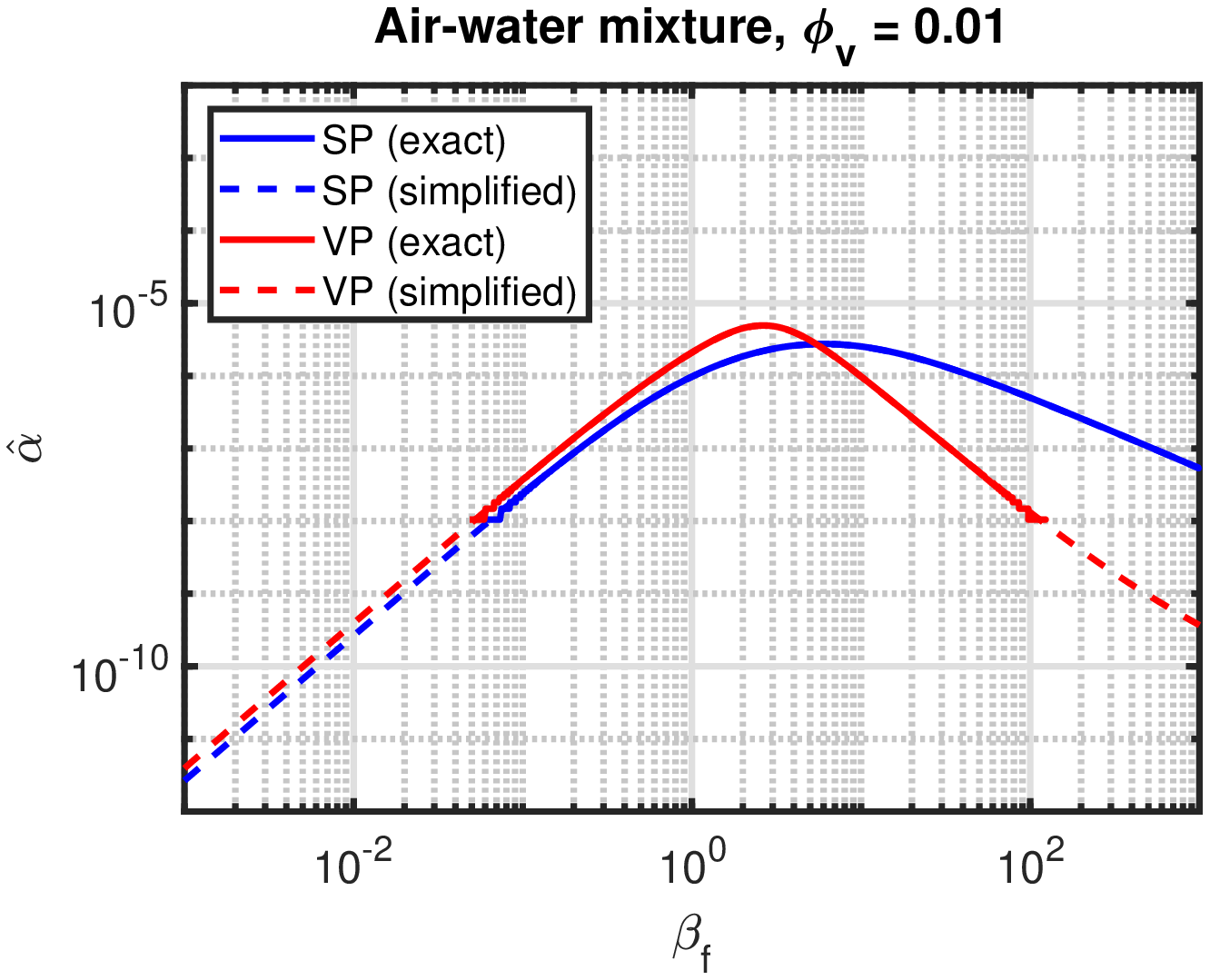}
\caption{Air-water mixture, left: Normalized sound speed, right: Nondimensional attenuation. Exact expressions are marked by solid lines and simplified expressions are marked by dashed lines.}
\label{fig:air_hydrosol}
\end{figure}

\begin{figure}
\centering
\includegraphics[width=6cm]{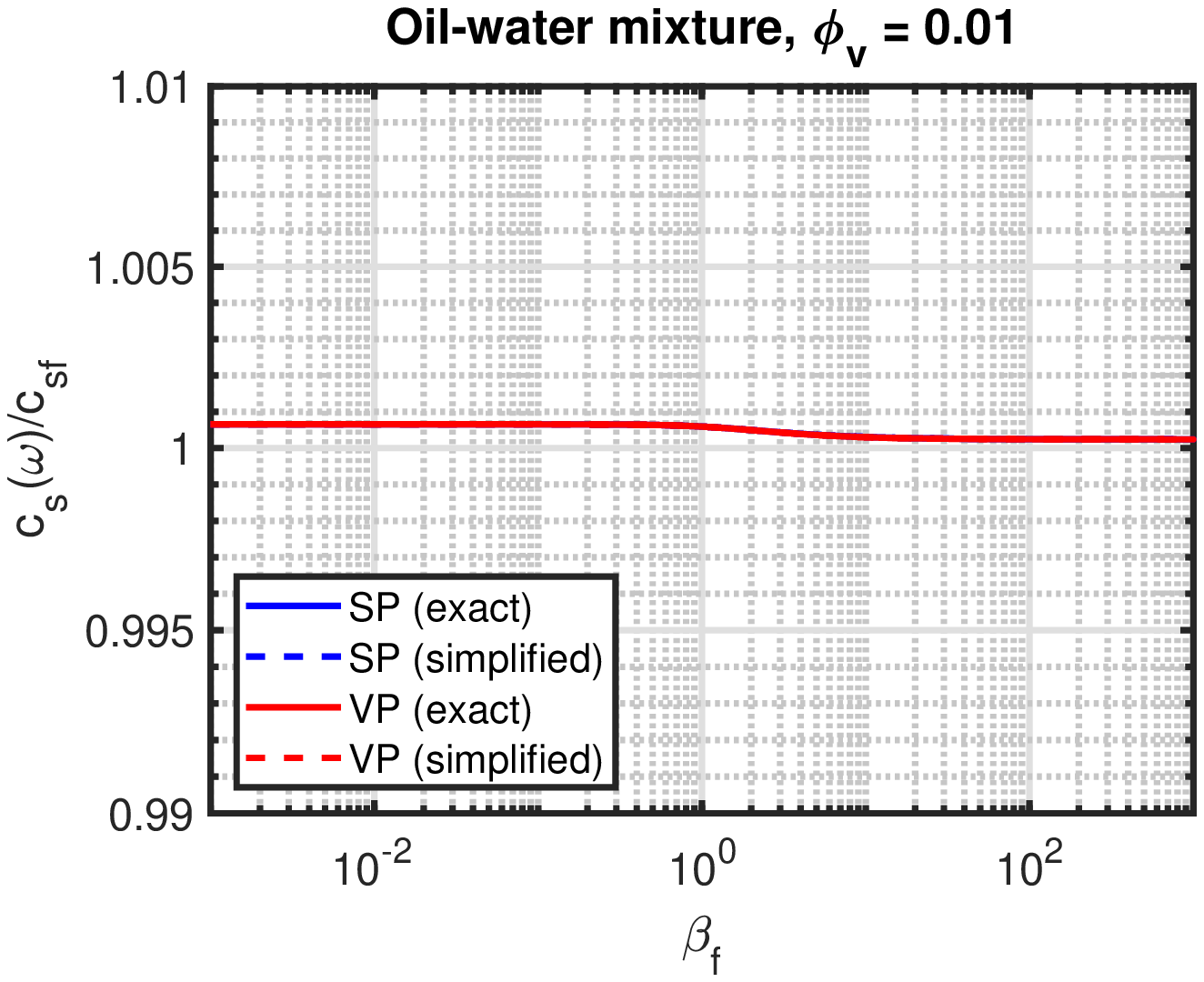}
\includegraphics[width=6cm]{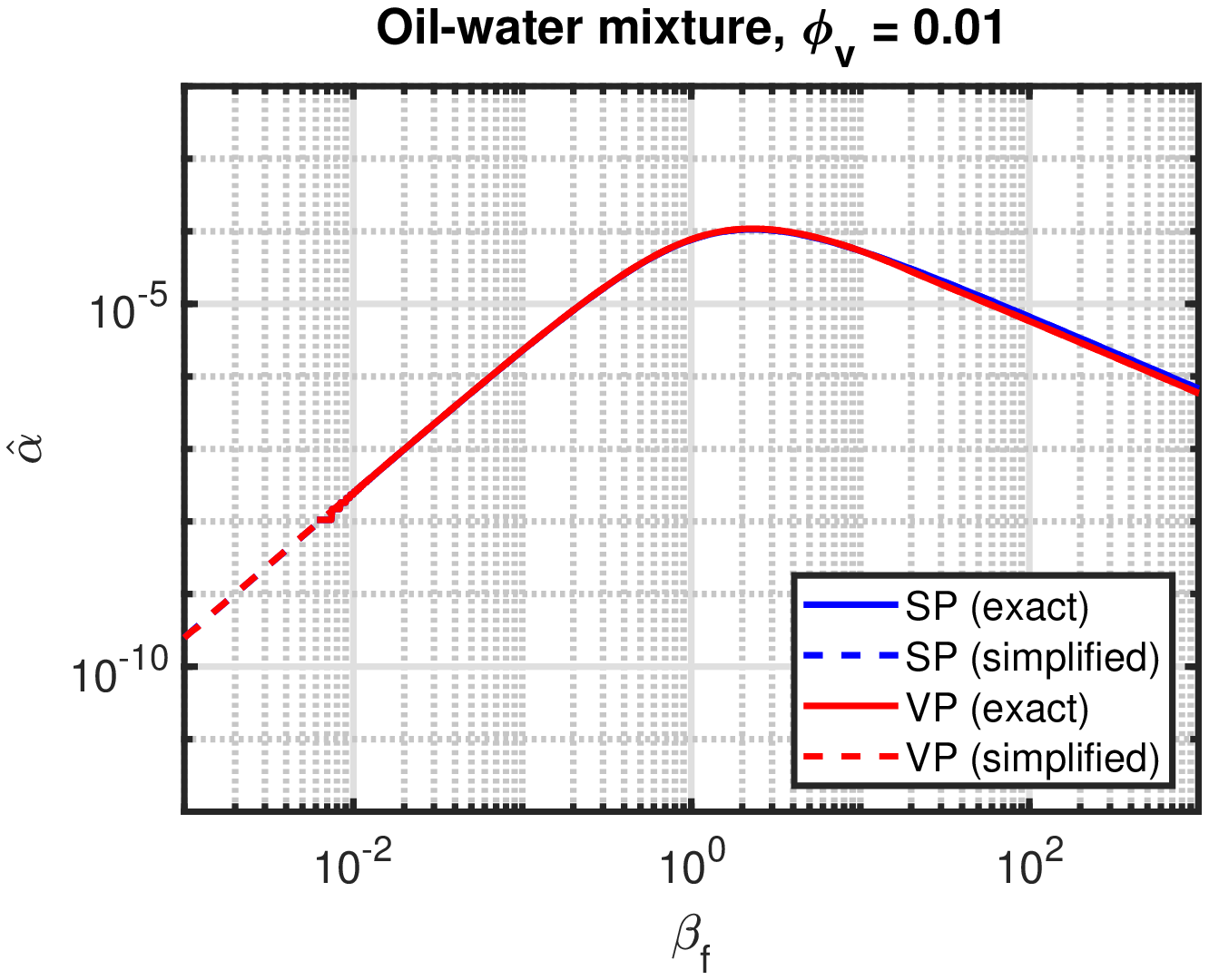}
\caption{Oil-water mixture, left: Normalized sound speed, right: Nondimensional attenuation. Exact expressions are marked by solid lines and simplified expressions are marked by dashed lines.}
\label{fig:oil_hydrosol}
\end{figure}

\begin{figure}
\centering
\includegraphics[width=6cm]{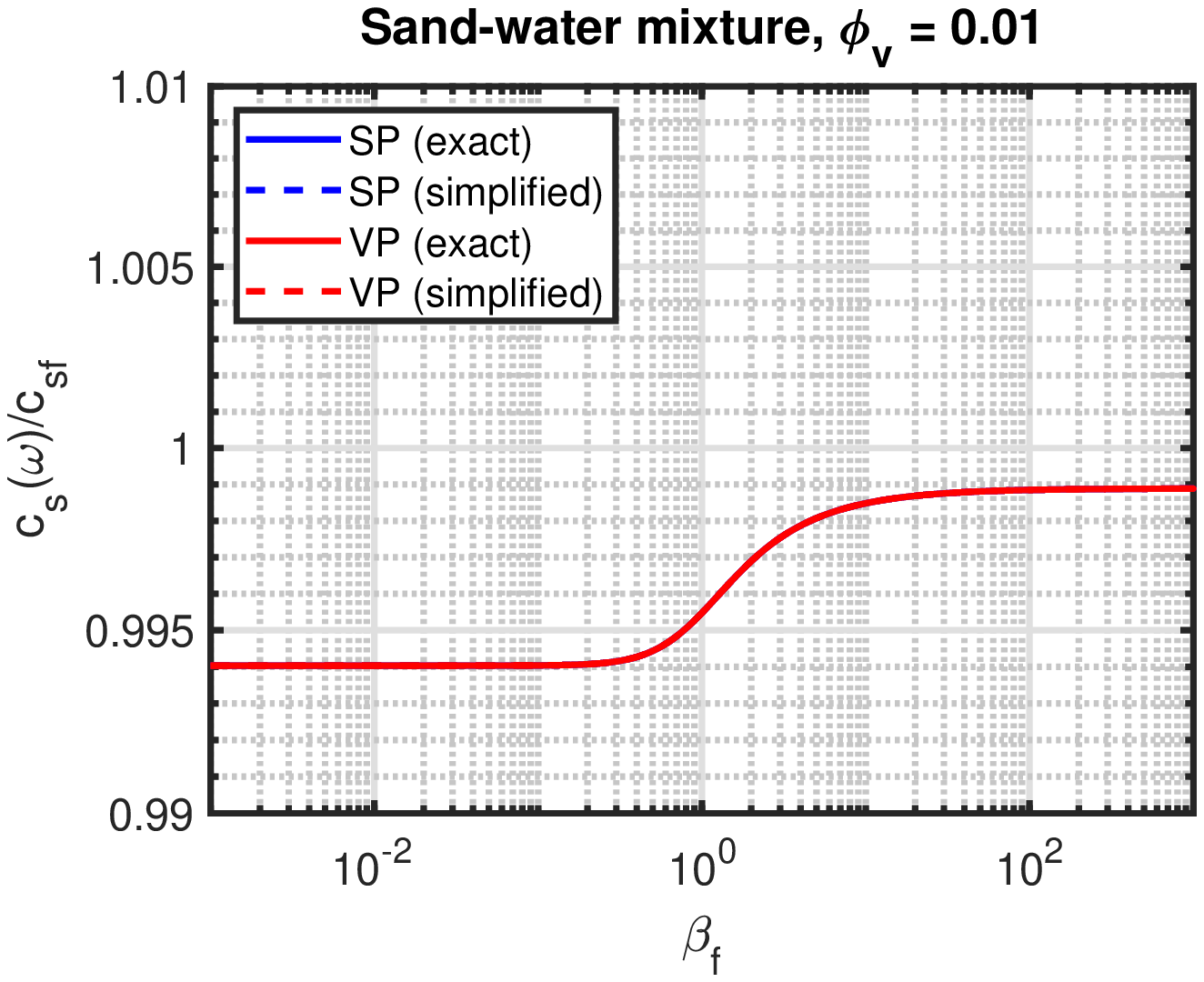}
\includegraphics[width=6cm]{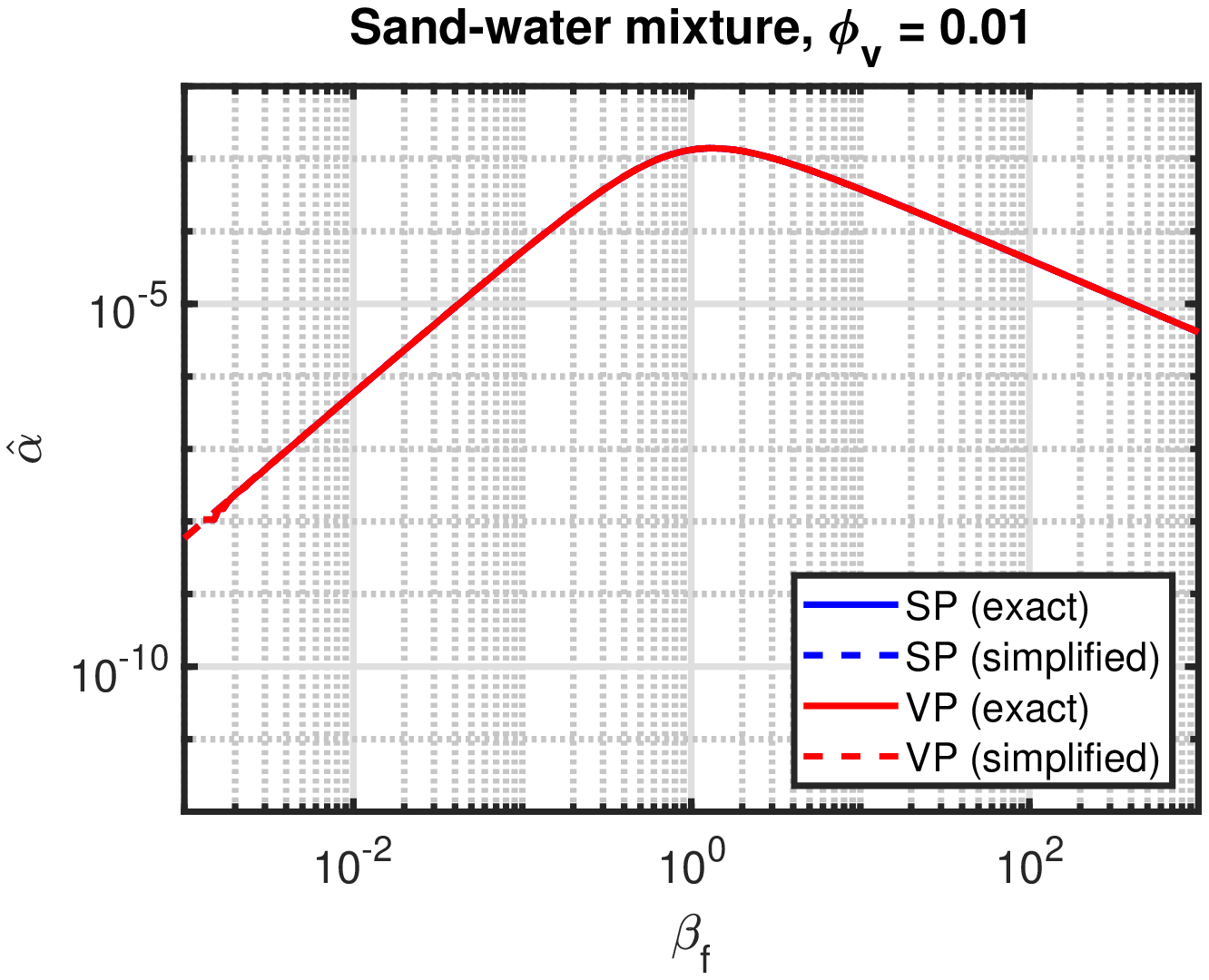}
\caption{Sand-water mixture, left: Normalized sound speed, right: Nondimensional attenuation. Exact expressions are marked by solid lines and simplified expressions are marked by dashed lines.}
\label{fig:sand_hydrosol}
\end{figure}

Only very small sound speed changes are occurring for the hydrosols as shown in Figures \ref{fig:air_hydrosol}-\ref{fig:sand_hydrosol}.

The peak nondimensional attenuation is several orders of magnitude below one, meaning that the simplified expression can be used instead of the exact expression.

We note that the curve for the exact expression for the nondimensional attenuation disappears for values below $10^{-8}$; this is because of roundoff errors when $Y/X$ approaches zero, see Equation (\ref{eq:alpha_hat}).

We observe that the nondimensional damping for the air-water mixture is different for the SP and VP theories; this corresponds to our findings for the amplitudes and phases of the velocity ratio, see Figure \ref{fig:amp_phase_hydrosols}.

For the oil-water and sand-water mixtures, the results from the SP and VP theories are almost identical.

\section{Conclusions}
\label{sec:conc}

We have extended the linear theory of isothermal sound propagation and attenuation in suspensions by applying the amended Coriolis flowmeter "bubble theory": Here, the drag force is a function of both the fluid and particle Stokes numbers and the particle-to-fluid ratio of the dynamic viscosity.

Aerosol and hydrosol examples are presented.

Our main result is that we have established a significant difference in damping between the theories for the air-water mixture, i.e. air bubbles entrained in water. Thus, for $\kappa = \mu_p/\mu_f \ll 1$, the dynamic viscosity of the particle modifies the attenuation significantly.

We have not found suitable, publicly available, measurements to compare to our findings: So we encourage the community to compare the extended theory to measurements.


\paragraph{Acknowledgements}

The author is grateful to Dr. John Hemp for creating, providing and explaining/discussing the Coriolis flowmeter bubble theory \cite{hemp_a}.




\end{document}